\documentclass[12pt]{article}
\pdfoutput =1
\usepackage{float} 
\usepackage{amsmath}
\usepackage{slashed}
\usepackage{amsfonts}   
\usepackage{amssymb}

\begin{document}
\date{}
\title{{\bf{\Large Newton-Cartan $ D 0$ branes from $ D1 $ branes and integrability}}}
\author{
 {\bf {\normalsize Dibakar Roychowdhury}$
$\thanks{E-mail:  dibakarphys@gmail.com, dibakar.roychowdhury@ph.iitr.ac.in}}\\
 {\normalsize  Department of Physics, Indian Institute of Technology Roorkee,}\\
  {\normalsize Roorkee 247667, Uttarakhand, India}
\\[0.3cm]
}

\maketitle
\begin{abstract}
We explore analytic integrability criteria for $ D1 $ branes probing 4D relativistic background with a null isometry direction. We use both the Kovacic's algorithm of classical (non)integrability as well as the standard formulation of Lax connections to show the analytic integrability of the associated dynamical configuration. We further use the notion of double null reduction and obtain the world-volume action corresponding to a torsional Newton-Cartan (TNC) $ D0 $ brane probing a 3D torsional Newton-Cartan geometry. Moreover, following Kovacic's method, we show the classical integrability of the TNC $ D0 $ brane configuration thus obtained. Finally, considering a trivial field redefinition for the $ D1 $ brane world-volume fields, we show the equivalence between two configurations in the presence of vanishing NS fluxes. 
\end{abstract}
\section{Overview and Motivation}
The extension of non relativistic (NR) string sigma models \cite{Gomis:2000bd}-\cite{Gomis:2005pg} to arbitrary backgrounds and understanding two of its primary aspects namely, (i) the UV completion and (ii) the underlying integrable structure (if any) stands extremely important in its own right. The target space geometry corresponding to NR propagating strings could be classified into two different categories. One of these goes under the name of string Newton-Cartan geometry obtained via gauging the centrally extended string Galilean algebra \cite{Bergshoeff:2019pij}-\cite{Gomis:2019zyu}. The other is obtained via null reduction of (relativistic) Lorentzian manifolds giving rise to what is known as torsional Newton-Cartan (TNC) geometry \cite{Harmark:2017rpg}-\cite{Roychowdhury:2020kma}. A recent analysis of \cite{Harmark:2019upf} reveals that under certain specific assumptions, these two seemingly different string theories could in principle be mapped into each other in a consistent manner.

The exciting evidence behind the existing integrable structure at the tree level of the Newton-Cartan (closed string) sigma models \cite{Roychowdhury:2019vzh},\cite{Roychowdhury:2019olt} has opened up a tremendous possibility of analyzing the NR stringy dynamics using the standard techniques of integrable models. This is therefore quite similar in spirit to that of its relativistic counterpart \cite{Beisert:2010jr}-\cite{Arutyunov:2009ga}. However, the understanding of similar questions in the corresponding open string sector still remains as a challange. The present article therefore aims to fill up some of these gaps and widen our current understanding beyond the closed string sector by taking into account the dynamics of extended objects like $ Dp $ branes\footnote{For our analysis, however, we stick to the special case with $ p=0 $.} \cite{Bachas:2000fr}-\cite{Petropoulos:2001qu} those probing the Galilean invariant manifolds. The corresponding target space geometry that we choose to work with happens to be a $ 2+1 $ dimensional TNC spacetime (those are obtained via null reduction of $ 3+1 $ dimensional Lorentzian manifolds \cite{Grosvenor:2017dfs}) with $ R \times S^2 $ topology. 

We start our analysis considering $ D1 $ branes propagating over 4D relativistic manifolds with a null isometry direction. Given the $ D1 $ brane configuration, we address the above issue of classical integrability following two traditional paths. One of these approaches goes under the name of Kovacic's algorithm \cite{kovacic1}-\cite{kovacic2} of classical (non)integrability which has been applied with remarkable success in various examples of relativistic sigma models \cite{Basu:2011fw}-\cite{Nunez:2018qcj} with or without supersymmetries. The other approach is based on the systematic formulation of Lax connection \cite{Arutyunov:2009ga} and thereby establishing its \emph{flatness} following the equations of motion. This further allows us to compute the infinite tower of conserved charges associated with the 2D world-volume theory and thereby proving the integrability. 

In the second part of the analysis, we use a double null reduction of the $ D1 $ brane world-volume action and obtain a world-volume description for torsional Newton-Cartan (TNC) $ D0 $ branes propagating over 3D torsional Newton-Cartan geometry with $ R\times S^2 $ topology. We further show the classical integrability of the configuration following Kovacic's method. On top of it, we show that following a trivial field redefinition, the $ D1 $ brane dynamics could be mapped to that of TNC $ D0 $ brane dynamics in the presence of vanishing NS fluxes. Finally, we draw our conclusion in Section 3.
\section{Road to integrability}
\subsection{Kovacic's method: A review}
For the sake of comprehensiveness, we briefly outline the essentials of Kovacic's algorithm that was proposed originally in \cite{kovacic1}. The algorithm essentially provides road to explore the classical (non)integrability criteria associated with dynamical phase space configurations. The steps are in fact quite straightforward to follow: (1) choose an invariant plane in the dynamical phase space and (2) consider fluctuations normal to this plane. These fluctuations generally obey differential equations,
\begin{eqnarray}
a(\tau)\ddot{\eta}(\tau)+b(\tau)\dot{\eta}(\tau)+c(\tau)\eta (\tau)=0
\label{e1}
\end{eqnarray}
known as Normal Variational Equations (NVEs) \cite{Stepanchuk:2012xi}. Here, $ a $ , $ b $ and $ c $ are in general complex \emph{rational} functions. The associated phase space configuration is said to be classically integrable if there exists simple algebraic/logarithmic/exponential solutions to (\ref{e1}) known as \emph{Liouvillian} solutions \cite{kovacic1}-\cite{kovacic2}, \cite{Basu:2011fw}-\cite{Stepanchuk:2012xi}. In summary, the algorithm sets rules to check whether NVE (\ref{e1}) admits Liouvillian solutions or not.

To check this explicitly, it is customary first to note down an equivalent representation \cite{kovacic1}-\cite{kovacic2} of (\ref{e1}),
\begin{eqnarray}
\ddot{\xi}=V(\tau)\xi (\tau)~;~V(\tau)=\frac{2\dot{b}a-2b\dot{a}+b^2 -4ac}{4a^2}.
\label{e2}
\end{eqnarray} 

Substituting, $\xi(\tau)\sim e^{\int w(\tau)d\tau} $ into (\ref{e2}) finally yields,
\begin{eqnarray}
\dot{w}(\tau)+w^2(\tau)=V(\tau)
\end{eqnarray}
where $ w(\tau) $ is a (complex) rational function of the form, $ \frac{P(\tau)}{Q(\tau)} $. Following the algorithm \cite{kovacic1}-\cite{kovacic2}, the NVE (\ref{e1}) allows Liouvillian form of solutions \emph{iff} $ w(\tau) $ turns out to be a polynomial of degree 1, 2, 4, 6 or 12. 

Interestingly enough, we discover that for extended objects like $ D1 $ branes (those probing 4D relativistic backgrounds) as well as nonrelativistic $ D0 $ branes (those probing 3D TNC geometries) it is indeed possible to find a very special form of NVEs (\ref{e1}) with $ a\neq 0 $ together with $ b=c=0 $ which therefore uniquely sets the potential $ V(\tau)=0 $ as well as the rational polynomial $ w(\tau)\sim \frac{1}{\tau} $ with degree 1. The most general expression for these Polynomials goes under the name of Mobius transformations that generate the group of automorphisms of the Riemann sphere.  
\subsection{Relativistic $ D1 $ branes}
We consider $ D1 $ brane dynamics over 4D relativistic backgrounds with a \emph{null} isometry direction. For technical simplicity, we set the dilaton as well as the background RR fluxes to zero and take into account only the background NS-NS fluxes ($ \mathcal{B}_{MN} $). 

The resulting DBI action \cite{Bachas:2000fr} is given by,
\begin{eqnarray}
\mathcal{S}_{Dp}=-T_{1}\int d^{2}\xi \sqrt{|\det \mathcal{A}_{\alpha \beta}|}\equiv -T_{1}\int d^{2}\xi ~ \mathcal{L}_{D1}
\label{en4}
\end{eqnarray}
where we identify,
\begin{eqnarray}
\mathcal{A}_{\alpha \beta}=\mathcal{G}_{MN}(X^P)\partial_{\alpha}X^{M}\partial_{\beta}X^N + l_{s}^2\mathcal{F}_{\alpha \beta}+\mathcal{B}_{MN}(X^P)\partial_{\alpha}X^M \partial_{\beta}X^N.
\label{e5}
\end{eqnarray}
Here, $ T_1 =l_{s}^{-2}$ stands for the $ D1 $ brane tension together with $ \xi^{\alpha}(\alpha =0,1) $ as world-volume directions. Moreover, we identify $ \mathcal{F}_{\alpha \beta}
=\partial_{\alpha}\mathfrak{a}_{\beta}-\partial_{\beta}\mathfrak{a}_{\alpha} $ as being the world-volume field strength tensor where $ \mathfrak{a}_{\alpha} $ is the corresponding $ U(1) $ gauge field. 

To proceed further, we consider the following 4D geometry,
\begin{eqnarray}
ds^2 = 2 \tau_{\mu}dX^{\mu}d\mathfrak{u}-2 \mathfrak{m}_{\varphi}\tau_{\mu}dX^{\mu}d\varphi +\frac{1}{4}(d \theta^2 + \sin^2 \theta d\varphi^2)
\label{e3}
\end{eqnarray}
where we identify individual metric functions \cite{Grosvenor:2017dfs},\cite{Roychowdhury:2019olt}
\begin{eqnarray}
\tau_{\mu}dX^{\mu}=\frac{1}{2}d\psi + dt -\frac{1}{2}\cos\theta d\varphi ~;~ \mathfrak{m}_{\varphi}=\frac{1}{4}\cos\theta.
\end{eqnarray}

Notice that, here $X^{\mathfrak{u}}\equiv \mathfrak{u} $ is the so called null isometry direction associated with the target space manifold. Using (\ref{e3}), it is therefore trivial to show
\begin{eqnarray}
\mathcal{A}_{\alpha \beta}=2 \partial_{\alpha}t\partial_{\beta}\mathfrak{u}-\frac{1}{2}\cos\theta \partial_{\alpha}t\partial_{\beta}\varphi +\partial_{\alpha}\psi \partial_{\beta}\mathfrak{u}-\cos\theta \partial_{\alpha}\varphi \partial_{\beta}\mathfrak{u}~~~~~~~~~~~~~~~~\nonumber\\
-\frac{1}{4}\cos\theta \partial_{\alpha}\psi \partial_{\beta}\varphi +\frac{1}{4}(\partial_{\alpha}\theta \partial_{\beta}\theta +\partial_{\alpha}\varphi \partial_{\beta}\varphi)+l_{s}^{2}\mathcal{F}_{\alpha \beta}+ \mathcal{B}\sin\theta \partial_{\alpha}\theta \partial_{\beta}\varphi
\end{eqnarray}
where we choose to work with NS-NS two form $\mathcal{B}_{\theta \varphi}=\mathcal{B}\sin\theta  $  \cite{Roychowdhury:2020kma} that corresponds to some specific values of the page charge, $ Q_D \sim \int_{S^2}\mathcal{B}_{2} $ which takes quantized values on the world-volume of the $ D1 $ brane. This is related to the underlying mechanism known as flux stabilization which states that the background NS fluxes sort of prevents $ D1 $ branes (wrapping $ S^2 $) from shrinking it to zero size.
\subsubsection{The world-volume theory}
To start with, we consider that the $ D1 $ brane is placed at a point $ X^{\mathfrak{u}}=$ constant, along the axis of null isometry. The ansatz that we choose to work with is that of a $ D1 $ brane wrapping the azimuthal direction of $ S^2 $,
\begin{eqnarray}
t = t(\tau) =\tau = \xi^0 ~;~ \theta = \theta (\tau)~;~ \psi = \psi (\tau)~;~\varphi = \kappa \sigma ~;~\mathfrak{a}_{\sigma}=\mathfrak{a}_{\sigma}(\tau)
\end{eqnarray}
where $ \kappa $ is the corresponding winding number.

The resulting matrix elements $  \mathcal{A}_{\alpha \beta}$  are given by,
\begin{eqnarray}
\mathcal{A}_{\tau \tau}&=&\frac{\dot{\theta}^2}{4} \\
\mathcal{A}_{\tau \sigma}&=&-\frac{\kappa}{2}\cos\theta (\dot{t}+\dot{\tilde{\psi}}) +l^2_s \dot{\mathfrak{a}}_{\sigma}+\kappa \mathcal{B}\dot{\theta}\sin\theta \\
\mathcal{A}_{\sigma \tau}&=&-l_{s}^{2}\dot{\mathfrak{a}}_{\sigma}\\
\mathcal{A}_{\sigma \sigma}&=& 0
\end{eqnarray}
where we define, $ \tilde{\psi}=\frac{\psi}{2} $.
\subsubsection{Equations of motion}
The corresponding Lagrangian density is given by,
\begin{eqnarray}
\mathcal{L}_{D1}=l_s \sqrt{\dot{\mathfrak{a}}_{\sigma}(-\frac{\kappa}{2}\cos\theta (\dot{t}+ \dot{\tilde{\psi}}) +l^2_s \dot{\mathfrak{a}}_{\sigma}+\kappa \mathcal{B}\dot{\theta}\sin\theta)}\equiv l_s \sqrt{\Gamma}
\end{eqnarray}
where we choose,
\begin{eqnarray}
\Gamma (\tau)=\dot{\mathfrak{a}}_{\sigma}(-\frac{\kappa}{2}\cos\theta (\dot{t}+ \dot{\tilde{\psi}}) +l^2_s \dot{\mathfrak{a}}_{\sigma}+\kappa \mathcal{B}\dot{\theta}\sin\theta).
\end{eqnarray}

The resulting equations of motion could be formally expressed as,
\begin{eqnarray}
\label{e37}
\Gamma(\kappa \dot{\mathfrak{a}}_{\sigma}\sin\theta (1+\dot{\tilde{\psi}})- 2 \mathcal{B}\kappa l_s \ddot{\mathfrak{a}}_{\sigma}\sin\theta ) + \kappa \mathcal{B}\dot{\mathfrak{a}}_{\sigma}\sin\theta \frac{d \Gamma}{d \tau} &=&0\\
\Gamma \kappa (\ddot{\mathfrak{a}}_{\sigma}\cos\theta - \dot{\mathfrak{a}}_{\sigma}\dot{\theta}\sin\theta) - \frac{\kappa}{2}\dot{\mathfrak{a}}_{\sigma}\cos\theta \frac{d \Gamma}{d \tau}& = & 0\\
\left(2 \Gamma \ddot{\mathfrak{a}}_{\sigma}-\dot{\mathfrak{a}}_{\sigma}\frac{d \Gamma}{d \tau} \right)(l^2_s \dot{\mathfrak{a}}_{\sigma}^2-\Gamma) &=&0.
\label{e39}
\end{eqnarray}
\subsubsection{Normal variational equations}
We set, $ \ddot{\mathfrak{a}}_{\sigma}=0 $ which yields the reduced set of equations,
\begin{eqnarray}
\label{e38}
\cos\theta \dot{\theta}(1+\dot{\tilde{\psi}}-\dot{\theta})+\sin\theta (\ddot{\tilde{\psi}}-\ddot{\theta})&=&0\\
\sin\theta \dot{\theta}(1+\dot{\tilde{\psi}})-\ddot{\tilde{\psi}}\cos\theta+2 \mathcal{B}(\ddot{\theta}\sin\theta + \dot{\theta}^2 \cos\theta)& =&0.
\label{e39}
\end{eqnarray}

Notice that, the second equation (\ref{e39}) is a direct consequence of setting $ \frac{d \Gamma}{d \tau}=0 $. In order to obtain NVE we set, $ \theta = \dot{\theta}=\ddot{\theta}=0 $ which trivially solves (\ref{e38}). This automatically fixes the corresponding invariant plane as $ \lbrace \theta=0 , \Pi_{\theta}=0 \rbrace$. Our aim would be solve fluctuations ($ \delta \theta(\tau) $) normal to this plane.

Substituting the above ansatz into (\ref{e39}) we find,
\begin{eqnarray}
\ddot{\tilde{\psi}}\Big|_{\theta \sim \Pi_{\theta}\sim 0} \sim \ddot{\psi}\Big|_{\theta \sim \Pi_{\theta}\sim 0}\approx 0
\end{eqnarray}
which thereby yields, $ \psi(\tau)\sim \tau + \mathfrak{c} $. Substituting this back into (\ref{e38}) and considering fluctuations $ \delta \theta\sim \eta(\tau) $ at leading order, we arrive at the following NVE
\begin{eqnarray}
\ddot{\eta}(\tau)\approx 0
\end{eqnarray}
which admits Loiuvillian solution of the form,
\begin{eqnarray}
\eta (\tau)\sim \tau + \mathfrak{C}.
\end{eqnarray}
Therefore, following our discussion in the previous Section, we conclude that the associated dynamical phase space configuration is classically integrable.
\subsubsection{Lax pairs and integrability}
In this Section, we look forward towards identifying the $ D1 $ brane dynamics in terms of a proper formulation of Lax connections \cite{Arutyunov:2009ga} over $ X^{\mathfrak{u}}= $ constant sub-manifold of the full relativistic/Lorentz invariant $ 3+1 $ dimensional manifold (\ref{e3}). The corresponding world-volume theory turns out to be, 
\begin{eqnarray}
\mathcal{S}_{D1}=-T_{1}\int d^{2}\xi \sqrt{|\det \mathcal{A}_{\alpha \beta}|}
\label{ee45}
\end{eqnarray}
where we identify,
\begin{eqnarray}
\mathcal{A}_{\alpha \beta}&=&-\frac{1}{2}\cos\theta \partial_{\alpha}v\partial_{\beta}\varphi 
+\frac{1}{4}(\partial_{\alpha}\theta \partial_{\beta}\theta +\partial_{\alpha}\varphi \partial_{\beta}\varphi)
+ \mathcal{B}\sin\theta \partial_{\alpha}\theta \partial_{\beta}\varphi +l_{s}^{2}\mathcal{F}_{\alpha \beta}\nonumber\\
&=& \mathcal{G}_{\alpha \beta}++l_{s}^{2}\mathcal{F}_{\alpha \beta}
\label{e46}
\end{eqnarray}
together with \cite{Grosvenor:2017dfs}, $ v=\frac{\psi}{2}+t $ which we collectively identify as time.
\subsubsection{The 2D world-volume current}
Our starting point is the consideration of the $ D1 $ brane dynamics over group manifold $ G\sim S^2$ with $ SO(3) $ isometries. The Killing generators that span the $\mathfrak{so}(3)\sim \mathfrak{su}(2) $ Lie algebra could be schematically expressed as \cite{Grosvenor:2017dfs},
\begin{eqnarray}
\mathfrak{T}_{a}=\mathfrak{e}_{a}~^{M}(X^{N})\partial_{M}~;~a=1, 2, 3
\label{e45}
\end{eqnarray} 
where $ \mathfrak{e}_{a}~^{M} $  $(X^M~;~M=v, \theta, \varphi) $ are the expansion coefficients that could be fit into the following  $ 3 \times 3 $ matrix as, 
\begin{equation}
[\mathfrak{e}]_{3 \times 3} = 
\begin{pmatrix}
\frac{\cos\varphi}{2 \sin\theta}&\sin\varphi & \cos\varphi \cot\theta \\ \\
\frac{-\sin\varphi}{2 \sin\theta}& \cos\varphi & -\sin\varphi \cot\theta \\ \\
0 &  0 & 1
\end{pmatrix}.
\end{equation}

In the following, we introduce 2D world-volume currents as
\begin{eqnarray}
\mathfrak{J}=g^{-1}dg\simeq \ell^{(M)}\mathfrak{e}^{a}~_{M} \mathfrak{T}_{a}dX^M~;~g\in G
\label{ee49}
\end{eqnarray}
subjected to the realization that $ \mathfrak{e}^{a}~_{M} $ are the elements of the inverse matrix $ [\mathfrak{e}^{-1}]_{3 \times 3} $ such that, $   \mathfrak{e}_b~^M\mathfrak{e}^{a}~_{M} = \delta^a_b $. 

An explicit computation further reveals,
\begin{equation}
[\mathfrak{e}^{-1}]_{3 \times 3} = \left(
\begin{array}{ccc}
 2 \cos \varphi  \sin \theta & -2 \sin \theta  \sin \varphi  & -2 \cos \theta  \\
 \sin \varphi  & \cos \varphi  & 0 \\
 0 & 0 & 1 \\
\end{array}
\right).
\end{equation}

To proceed further, next we note down
\begin{eqnarray}
d \mathfrak{J}=d(g^{-1}dg)=-g^{-2}dg \wedge dg = -\mathfrak{J}\wedge \mathfrak{J}
\label{ee50}
\end{eqnarray}
which thereby yields the identity of the following form,
\begin{eqnarray}
d \mathfrak{J}+\mathfrak{J}\wedge \mathfrak{J}=0.
\label{ee51}
\end{eqnarray}

Substituting (\ref{ee49}) into (\ref{ee51}) we find,
\begin{eqnarray}
\partial_N \tilde{\mathfrak{e}}^{a}~_{M}-\partial_M \tilde{ \mathfrak{e}}^{a}~_{N}+\tilde{\mathfrak{e}}^{b}~_{M}\tilde{\mathfrak{e}}^{c}~_{N}\epsilon_{bc}~^{a}=0
\label{ee53}
\end{eqnarray}
where, $ \epsilon_{abc} $ is the structure constant of the underlying $ \mathfrak{so}(3) $ Lie algebra. Moreover, we absorb the proportionality constant into the definition of $ \mathfrak{e}^{a}~_{M} $ namely, $ \tilde{\mathfrak{e}}^a~_{M}=\ell^{(M)} \mathfrak{e}^a~_{M}$.

On the other hand, expressing the current explicitly into its components we find,
\begin{eqnarray}
\mathfrak{J}_{\alpha}&=& \mathfrak{J}_{\alpha}^{a} \mathfrak{T}_{a}\\
\mathfrak{J}_{\alpha}^{a}&=&\ell^{(i)}\mathfrak{e}^{a}~_{i}\partial_{\alpha}X^{i}+\ell^{(\varphi)}\mathfrak{e}^{a}~_{\varphi}\partial_{\alpha}\varphi ~;~i=v ,\theta.
\label{e55}
\end{eqnarray}

Using (\ref{e55}), it is infact trivial to show
\begin{eqnarray}
\mathfrak{J}_{\alpha}^{a}~\mathfrak{J}_{\beta}^{b}~\Omega_{ab}=\ell^{(i)}\ell^{(j)}\mathfrak{e}^{a}~_{i}\mathfrak{e}^{b}~_{j}\Omega_{ab}\partial_{\alpha}X^i \partial_{\beta}X^j +\ell^{(i)}\ell^{(\varphi)}\mathfrak{e}^{a}~_{i}\mathfrak{e}^{b}~_{\varphi}\Omega_{ab}\partial_{\alpha}X^i \partial_{\beta}\varphi 
\nonumber\\
+\ell^{(\varphi)}\ell^{(i)}\mathfrak{e}^{a}~_{\varphi}\mathfrak{e}^{b}~_{i}\Omega_{ab}\partial_{\alpha}\varphi \partial_{\beta}X^i +(\ell^{(\varphi)})^2\mathfrak{e}^{a}~_{\varphi}\mathfrak{e}^{b}~_{\varphi}\Omega_{ab}\partial_{\alpha}\varphi \partial_{\beta}\varphi.
\label{e51}
\end{eqnarray}

In order to map (\ref{e51}) into (\ref{e46}) we impose the following set of constraints,
\begin{eqnarray}
\label{e52}
\mathfrak{e}^{a}~_{i}\mathfrak{e}^{b}~_{j}\Omega_{ab}&=&\delta_{i}^{\theta}\delta_{j}^{\theta}~;~\mathfrak{e}^{a}~_{\varphi}\mathfrak{e}^{b}~_{i}\Omega_{ab}=0\\
\mathfrak{e}^{a}~_{\varphi}\mathfrak{e}^{b}~_{\varphi}\Omega_{ab}&=&1~;~\mathfrak{e}^{a}~_{v}\mathfrak{e}^{b}~_{\varphi}\Omega_{ab}=-\cos\theta\\
\mathfrak{e}^{a}~_{\theta}\mathfrak{e}^{b}~_{\varphi}\Omega_{ab}&=&4\mathcal{B}\sin\theta
\label{e54}
\end{eqnarray}
together with the coefficients, $ \ell^{(\theta)}=\ell^{(\varphi)}=\frac{1}{2} $ and $ \ell^{(v)}=1 $. 

The above set of equations (\ref{e52})-(\ref{e54}) could be decomposed into a set of six linear algebraic equation in $ \Omega_{ab} (\theta , \varphi)$ which in principle can be solved for nine different elements in $ \Omega_{3 \times 3} $. Below we enumerate the set of solutions satisfying (\ref{e52})-(\ref{e54}),
\begin{eqnarray}
\Omega_{11}&=&\frac{1}{4\sin^2\theta \sin^2\varphi}~;~\Omega_{22}=\frac{1}{\cos^2 \varphi}\\
\Omega_{12}&=&\Omega_{21}=\frac{1}{4 \sin\theta \sin\varphi \cos\varphi}\\
\Omega_{32}&=&\frac{\cot\theta}{2\sin\varphi \cos\varphi}~;~\Omega_{31}=\frac{\cot\theta}{2\sin\theta \sin^2\varphi}\\
\Omega_{23}&=&- \sin\varphi \cos\theta + 4 \mathcal{B}\cos\varphi \sin\theta +\frac{\cot \theta}{2 \sin\varphi \cos\varphi}\\
\Omega_{13}&=&- \frac{1}{2}\cos\varphi \cot\theta -2 \mathcal{B}\sin\varphi +\frac{\cos \theta}{2 \sin^2\varphi \sin^2 \theta}\\
\Omega_{33}&=&1+2\cos\theta \Omega_{13}.
\end{eqnarray}
\subsubsection{Equations of motion}
With the above set-up in hand, the $ D1 $ world-volume theory (\ref{ee45}) could be formally expressed as,
\begin{eqnarray}
\mathcal{S}_{D1}&=&-T_{1}\int d^{2}\xi \sqrt{|\det \mathcal{A}_{\alpha \beta}|}\\
\mathcal{A}_{\alpha \beta}&=& \mathfrak{J}_{\alpha}^{a}~\mathfrak{J}_{\beta}^{b}~\Omega_{ab}(\theta , \varphi)+l^2_s \mathcal{F}_{\alpha \beta}\nonumber\\
&=&  \mathfrak{J}_{\alpha}\cdot  \mathfrak{J}_{\beta}+l^2_s \mathcal{F}_{\alpha \beta}
\end{eqnarray}
where we introduce the notation, $ \mathfrak{J}_{\alpha}\cdot \mathfrak{J}_{\beta}\equiv\mathfrak{J}_{\alpha}^{a}~\mathfrak{J}_{\beta a}=  \mathfrak{J}_{\alpha}^{a}~\mathfrak{J}_{\beta}^{b}~\Omega_{ab}$.

Before we proceed further, it is customary first to note down the following identity
\begin{eqnarray}
\delta \mathfrak{J}^{a}_{\alpha} &=& (\partial_M \tilde{ \mathfrak{e}}^{a}~_{N}-\partial_N \tilde{\mathfrak{e}}^{a}~_{M})\partial_{\alpha}X^{N}\delta X^{M}+\partial_{\alpha}\delta \mathfrak{J}^a\nonumber\\
& = & (\tilde{\mathfrak{e}}^{b}~_{M}\tilde{\mathfrak{e}}^{c}~_{N}\epsilon_{bc}~^{a}\partial_{\alpha}X^{N})\tilde{\mathfrak{e}}_m~^{M}\delta \mathfrak{J}^m+\partial_{\alpha}\delta \mathfrak{J}^a
\label{ee68}
\end{eqnarray} 
where we use the fact, $\delta \mathfrak{J}^a = \tilde{\mathfrak{e}}^a~_{M}\delta X^M  $.

Using (\ref{ee68}), the equation of motion corresponding to world-volume currents could be formally expressed as,
\begin{eqnarray}
\partial_{\alpha}\left( \sqrt{|\det \mathcal{A}_{\alpha \beta}| }~ \mathfrak{g}^{\alpha \beta}~\mathfrak{J}_{\beta a}\right) =0
\label{ee69}
\end{eqnarray}
where we identify, $  \mathfrak{g}^{\alpha \beta}=\mathcal{A}^{\alpha \beta}+\mathcal{A}^{\beta \alpha} = \mathfrak{g}^{ \beta \alpha}$ as a symmetric tensor filed on the world-volume of the $ D1 $ brane. Moreover, here $ \mathcal{A}^{\alpha \beta} $ has been introduced as an inverse of the induced world-volume metric namely, $ \mathcal{A}^{\alpha \beta}\mathcal{A}_{\beta \gamma}=\delta^{\alpha}_{\gamma} $. 

On the other hand, the dynamics associated to world-volume gauge fields reveals,
\begin{eqnarray}
\label{ee70}
\partial_{\tau}\left(\frac{\mathfrak{b}_{\tau \sigma}+l^2_s \mathcal{F}_{\tau \sigma}}{\sqrt{|\det \mathcal{A}_{\alpha \beta}| }} \right)  &=&0\\
\partial_{\sigma}\left(\frac{\mathfrak{b}_{\tau \sigma}+l^2_s \mathcal{F}_{\tau \sigma}}{\sqrt{|\det \mathcal{A}_{\alpha \beta}| }} \right)  &=&0
\label{ee71}
\end{eqnarray}
where we introduce, $ \mathfrak{b}_{\tau \sigma}=\frac{1}{2}(\mathfrak{J}^{T}_{\tau}~\Omega~ \mathfrak{J}_{\sigma}-\mathfrak{J}^{T}_{\sigma}~\Omega~ \mathfrak{J}_{\tau}) =-\mathfrak{b}_{\sigma \tau}$. 

The above set of equations (\ref{ee70})-(\ref{ee71}) could be combined into a single equation,
\begin{eqnarray}
\varepsilon^{\alpha \beta}\partial_{\beta}\varpi_{\alpha \lambda}=0
\end{eqnarray}
where, $ \varpi_{\alpha \beta}=\frac{\mathfrak{b}_{\alpha \beta}+l^2_s \mathcal{F}_{\alpha \beta}}{\sqrt{|\det \mathcal{A}_{\alpha \beta}| }} $ is the \emph{antisymmetric} two form on the world-volume. Furthermore, as a natural consequence of (\ref{ee70})-(\ref{ee71}) it is in fact trivial to see, $ \varpi_{\tau \sigma}=\Pi= $ constant.
\subsubsection{The flat connection}
Given the above dynamics, we propose the Lax connection of the following form
\begin{eqnarray}
\mathfrak{L}^{a}_{\alpha}=\wp_{1}~\mathfrak{J}^a_{\alpha}+\wp_2 ~ \varepsilon_{\alpha \gamma}\sqrt{|\det \mathcal{A}_{\alpha \beta}| } ~ \mathfrak{g}^{\gamma \lambda}~\mathfrak{J}_{\lambda b}~ \delta^{b a}+\wp_3 ~\varpi_{\alpha \gamma}~\varepsilon^{\gamma \lambda}~\mathfrak{J}^{a}_ \lambda
\label{ee73}
\end{eqnarray}
where, $ \wp_{1} $, $ \wp_2 $ and $ \wp_3 $ are arbitrary constats that will be fixed from the flatness condition \cite{Arutyunov:2009ga} of the Lax connection. Moreover, here $ \varepsilon^{\tau \sigma}=-\varepsilon^{\sigma \tau}=1 $ is the 2D Levi-Civita symbol together with its inverse $ \varepsilon_{\alpha \beta}(=-\varepsilon_{\beta \alpha}) $ which is defined through the relation, $ \varepsilon^{\alpha \beta}\varepsilon_{\alpha \gamma}=\delta^{\beta}_{\gamma} $.

Using (\ref{ee53}), (\ref{ee69}), (\ref{ee70}) and (\ref{ee71}) it is now quite straightforward to show,
\begin{eqnarray}
\varepsilon^{\alpha \beta}\partial_{\beta}\mathfrak{L}^{a}_{\alpha}&=&-\partial_{\tau}\mathfrak{L}^{a}_{\sigma}+\partial_{\sigma}\mathfrak{L}^{a}_{\tau}\nonumber\\
&=&(\wp_1 - \wp_3 \Pi) \tilde{\mathfrak{e}}^{b}~_{M}\tilde{\mathfrak{e}}^{c}~_{N}\epsilon_{bc}~^{a}\partial_{\tau}X^N \partial_{\sigma}X^M.
\label{ee74}
\end{eqnarray}

On the other hand, after some trivial algebra we find
\begin{eqnarray}
\mathfrak{L}^{b}_{\tau}\mathfrak{L}^{c}_{\sigma}\epsilon_{bc}~^a = -(\wp_1 -\wp_3 \Pi)^2 ~\tilde{\mathfrak{e}}^{b}~_{M}\tilde{\mathfrak{e}}^{c}~_{N}\epsilon_{bc}~^{a}\partial_{\tau}X^N \partial_{\sigma}X^M~~~\nonumber\\
-8\wp_2^2 ~\tilde{\mathfrak{e}}^{b}~_{M}\tilde{\mathfrak{e}}^{c}~_{N}\epsilon_{bc}~^{a}\partial_{\tau}X^N \partial_{\sigma}X^M.
\label{ee75}
\end{eqnarray}

Combing (\ref{ee74}) and (\ref{ee75}) together, we arrive at the following relation
\begin{eqnarray}
\partial_{\tau}\mathfrak{L}^{a}_{\sigma}-\partial_{\sigma}\mathfrak{L}^{a}_{\tau}-\mathfrak{L}^{b}_{\tau}\mathfrak{L}^{c}_{\sigma}\epsilon_{bc}~^a=(\wp_1 - \wp_3 \Pi)(\wp_1 - \wp_3 \Pi -1)\tilde{\mathfrak{e}}^{b}~_{M}\tilde{\mathfrak{e}}^{c}~_{N}\epsilon_{bc}~^{a}\partial_{\tau}X^N \partial_{\sigma}X^M\nonumber\\
+8\wp_2^2 ~\tilde{\mathfrak{e}}^{b}~_{M}\tilde{\mathfrak{e}}^{c}~_{N}\epsilon_{bc}~^{a}\partial_{\tau}X^N \partial_{\sigma}X^M.
\label{ee76}
\end{eqnarray}

The flatness condition \cite{Arutyunov:2009ga} of Lax implies that the R.H.S. of (\ref{ee76}) must vanish identically. This is naturally achieved by setting the following constraint,
\begin{eqnarray}
(\wp_1 - \wp_3 \Pi)(\wp_1 - \wp_3 \Pi -1)+8 \wp_2^2 = 0.
\label{ee77}
\end{eqnarray}

A non trivial choice that solves (\ref{ee77}) could be of the form,
\begin{eqnarray}
\wp_1 = \wp_3 =\frac{1}{1-\Pi^2 }~;~ \wp_2 = \frac{\sqrt{\Pi}}{2\sqrt{2}(1+\Pi)}.
\label{ee78}
\end{eqnarray}

Using (\ref{ee78}), the flat connection (\ref{ee73}) could be formally expressed as,
\begin{eqnarray}
\mathfrak{L}^{a}_{\alpha}=\frac{1}{ 1 - \Pi^2 }\left(\mathfrak{J}^a_{\alpha}+\varpi_{\alpha \gamma}~\varepsilon^{\gamma \lambda}~\mathfrak{J}^{a}_ \lambda + \frac{\sqrt{\Pi}(1- \Pi)}{2\sqrt{2}} \varepsilon_{\alpha \gamma}\sqrt{|\det \mathcal{A}_{\alpha \beta}| } ~ \mathfrak{g}^{\gamma \lambda}~\mathfrak{J}_{\lambda }^b~ \delta^{a}_b\right) 
\label{ee79}
\end{eqnarray}
where, we identify $ \Pi $ as the \emph{spectral} parameter associated with the 2D integrable model. 
\subsubsection{Conserved charges}
The flat connection (\ref{ee79}) estimated above could be used to define tower of conserved charges associated with the 2D world-volume theory. The first step is to introduce the so called \emph{monodromy} matrix \cite{Beisert:2010jr}-\cite{Arutyunov:2009ga},
\begin{eqnarray}
\mathcal{T}(\Pi)=\mathcal{P}\exp \int_{0}^{2\pi}d \sigma ~ \mathfrak{L}_{\sigma}(\Pi):= \exp \Im (\Pi)
\label{ee80}
\end{eqnarray}
where we presume that one of the world-volume directions of $ D1 $ brane is compact where the world-volume fields obey periodic boundary conditions.

Using (\ref{ee80}), it is in fact straightforward to show,
\begin{eqnarray}
\partial_{\tau}\mathcal{T}=[\mathfrak{L}_{\tau},\mathcal{T}(\Pi)].
\label{ee81}
\end{eqnarray}

The next step would be to introduce the \emph{transfer} matrix,
\begin{eqnarray}
\mathsf{T} (\Pi)= tr\mathcal{T}(\Pi) =\sum_{n=0}^{\infty}\mathcal{Q}_n ~ \Pi^{n}
\end{eqnarray}
which by means of (\ref{ee81}) yields,
\begin{eqnarray}
\partial_{\tau}\mathcal{Q}_{n}=0~;~\forall n
\end{eqnarray}
an infinite tower of conserved charges,
\begin{eqnarray}
\mathcal{Q}_{n} =\frac{1}{n !}\frac{\partial^{n}}{\partial \Pi^{n}}~tr \sum_{m,k=0}^{\infty}\frac{1}{k !}\left(\int_{0}^{2 \pi}d\sigma\frac{\Pi^{m}}{m!}\frac{\partial^{m}}{\partial \Pi^{m}}\mathfrak{L}_{\sigma} (\Pi)\right)^{k}\Big|_{\Pi =0}
\end{eqnarray}
associated with the $ D1 $ brane dynamics over 4D relativistic backgrounds.
\subsection{Newton-Cartan $ D0 $ branes}
We now explore integrability criteria for Newton-Cartan (NC) $ D0 $ branes (those probing 3D TNC geometries with $ R \times S^2 $ topology) starting from the world-volume action of the 4D $ D1 $ branes (\ref{en4}) propagating over the relativistic background. This is achieved following a double null reduction, 
 \begin{eqnarray}
 \mathfrak{u}=\xi^{1}=\sigma
 \end{eqnarray}
where rest of the world-volume d.o.f. are taken to be independent of $ \xi^{1} $.

Below we enumerate different elements of $ 2 \times 2 $ matrix $ \mathcal{A}_{\alpha \beta} $,
\begin{eqnarray}
\label{e7}
\mathcal{A}_{\tau \tau}&=&-\frac{1}{2}\cos\theta \dot{t}\dot{\varphi}-\frac{1}{4}\cos\theta \dot{\psi}\dot{\varphi}+\frac{1}{4}(\dot{\theta}^2 + \dot{\varphi}^2)+\mathcal{B}\sin\theta \dot{\theta}\dot{\varphi}\\
\mathcal{A}_{\tau \sigma}&=&2\dot{t}+\dot{\psi}-\cos\theta \dot{\varphi}+l_{s}^{2}\dot{\mathfrak{a}}_{\sigma} \\
\mathcal{A}_{\sigma \tau}&=&-l_{s}^{2}\dot{\mathfrak{a}}_{\sigma}\\
\mathcal{A}_{\sigma \sigma}&=& 0.
\label{e10}
\end{eqnarray}
\subsubsection{Conserved charges}
Given (\ref{e7})-(\ref{e10}), the corresponding Lagrangian density is given by,
\begin{eqnarray}
\mathcal{L}_{D0}=l_s\sqrt{\dot{\mathfrak{a}}_{\sigma}(2 \dot{t}+\dot{\psi}-\cos\theta \dot{\varphi}+l_{s}^{2}\dot{\mathfrak{a}}_{\sigma})}\equiv l_s \sqrt{\Lambda}
\end{eqnarray}
where we set, 
\begin{eqnarray}
 \Lambda (\tau)= \dot{\mathfrak{a}}_{\sigma}(2 \dot{t}+\dot{\psi}-\cos\theta \dot{\varphi}+l_{s}^{2}\dot{\mathfrak{a}}_{\sigma}).
\end{eqnarray}

The corresponding (conserved) charge densities associated with the $ D0 $ brane configuration are given by,
\begin{eqnarray}
\mathcal{E} &=&  \frac{\delta \mathcal{L}_{D0}}{\delta \dot{t}}=\frac{l_s \dot{\mathfrak{a}}_{\sigma}}{\sqrt{\dot{\mathfrak{a}}_{\sigma}(2 \dot{t}+\dot{\psi}-\cos\theta \dot{\varphi}+l_{s}^{2}\dot{\mathfrak{a}}_{\sigma})}}\\
\Pi_{\varphi}& =&\frac{\delta \mathcal{L}_{D0}}{\delta \dot{\varphi}}=\frac{-l_s \dot{\mathfrak{a}}_{\sigma}\cos\theta}{\sqrt{\dot{\mathfrak{a}}_{\sigma}(2 \dot{t}+\dot{\psi}-\cos\theta \dot{\varphi}+l_{s}^{2}\dot{\mathfrak{a}}_{\sigma})}}.
\end{eqnarray}
\subsubsection{Equations of motion}
Next, we note down equations of corresponding to different world-volume fields. To proceed further, we set $ t =\tau = \xi^0$. Below we enumerate equations of motion for world volume fields,
\begin{eqnarray}
\label{e14}
\dot{\varphi}\dot{\mathfrak{a}}_{\sigma}\sin\theta &=&0\\
\label{e16}
 2 \Lambda \ddot{\mathfrak{a}}_{\sigma}-\dot{\mathfrak{a}}_{\sigma}\frac{d \Lambda}{d \tau}& =&0\\
\left(2 \Lambda \ddot{\mathfrak{a}}_{\sigma}-\dot{\mathfrak{a}}_{\sigma}\frac{d \Lambda}{d \tau} \right) \cos\theta -2 l^2_s \Lambda \dot{\mathfrak{a}}_{\sigma}\dot{\theta}\sin\theta & = & 0\\
\left(2 \Lambda \ddot{\mathfrak{a}}_{\sigma}-\dot{\mathfrak{a}}_{\sigma}\frac{d \Lambda}{d \tau} \right)(l^2_s \dot{\mathfrak{a}}_{\sigma}^2-\Lambda)& =&0.
\label{e17}
\end{eqnarray}
\subsubsection{Normal variational equations}
The above set of equations (\ref{e14})-(\ref{e17}) are essentially the basic ingredients of what we call Normal Variational Equations (NVEs) \cite{Basu:2011fw}-\cite{Stepanchuk:2012xi}. To start with, we choose to work with the dynamical phase space configuration with $\mathcal{F}_{\tau \sigma}= \dot{\mathfrak{a}}_{\sigma}\neq 0 $ (= \emph{constant}) together with $ \psi =$ constant. Notice that, both the \emph{invariant} planes that we choose below belong to the $ \psi = $ constant and $ \Pi_{\psi}= $ constant subspace of the full dynamical phase space configuration.  This further simplifies (\ref{e14})-(\ref{e17}),
\begin{eqnarray}
\label{e20}
\dot{\varphi}\dot{\theta}\cos\theta + \ddot{\varphi}\sin\theta -2l^2_s\Lambda \ddot{\theta}\sin\theta - 2l^2_s\Lambda \dot{\theta}^2\cos\theta &=& 0\\
\sin\theta \dot{\theta}\dot{\varphi}-\cos\theta\ddot{\varphi} & =& 0
\label{e21}
\end{eqnarray} 
where the second equation (\ref{e21}) follows from setting $ \frac{d \Lambda}{d \tau}=0 $.

In order to implement Kovacic's algorithm, we further choose to work with the following invariant plane $\theta= \dot{\theta}=\ddot{\theta}=0 $  which identically satisfies (\ref{e20}). The invariant plane one might wish to think of as a submanifold with, $ \theta =0 $ , $ \Pi_{\theta}=0 $ and $ \Pi_{\varphi}= $ constant within the subspace of the full dynamical phase space configuration. 

Upon substitution into (\ref{e21}) this further yields,
\begin{eqnarray}
\ddot{\varphi}\Big|_{\theta \sim \dot{\theta}\sim 0}\approx 0
\end{eqnarray}
which has a solution, 
\begin{eqnarray}
\varphi (\tau) \approx 2 l^2_s \Lambda~ \tau.
\label{e23}
\end{eqnarray}

Substituting (\ref{e23}) into (\ref{e20}) and considering infinitesimal fluctuations $ \delta\theta (\tau) \sim \eta (\tau) $ normal to the invariant plane, we arrive at the following NVE
\begin{eqnarray}
\ddot{\eta}(\tau)\approx 0
\label{ee26}
\end{eqnarray}
where we retain ourselves upto leading order in the fluctuations ($ \eta (\tau) $) and also take into account of the fact $\Big| \frac{\delta \eta}{\eta}\Big|\ll1 $. Under the above set of assumptions, the NVE (\ref{ee26}) allows \emph{Liouvillian} solution of the following form,
\begin{eqnarray}
\eta (\tau)\sim   \tau + \mathfrak{c}
\end{eqnarray}
which ensures the integrability of the associated phase space configuration. 

The second phase space configuration that one might choose to work with is to set the invariant plane as, $\varphi =  \dot{\varphi}=\ddot{\varphi}=0 $ which trivially solves (\ref{e21}). This is a submanifold that satisfies, $ \varphi =0 $ and $ \Pi_{\varphi}= 0$. Substituting this into (\ref{e20}) yields,
\begin{eqnarray}
\dot{\theta}\sin\theta \Big|_{\varphi \sim \dot{\varphi}\sim  0}\approx 0
\label{e26}
\end{eqnarray}
which thereby sets, $ \theta (\tau) = \theta_c =$ constant. Substituting (\ref{e26}) into (\ref{e21}) and considering fluctuations $ \delta\varphi (\tau)\sim \bar{\eta}(\tau) $ normal to the invariant plane in the phase space we find,
\begin{eqnarray}
\ddot{\bar{\eta}}(\tau)\approx 0
\end{eqnarray}
which admits the Liouvillian solution of the following form,
\begin{eqnarray}
\bar{\eta}(\tau)\sim \tau + \bar{\mathfrak{c}}.
\end{eqnarray}
The above analysis confirms that the second phase space configuration is also classically integrable in the sense of Kovacic.
\subsection{Mapping from $ D1$ to $ D0 $ }
We now show that under certain specific assumptions the effective $ D1 $ brane dynamics (\ref{e38})-(\ref{e39}) reduces to that of TNC $ D0 $ brane dynamics as found above in (\ref{e20})-(\ref{e21}). In order to show the equivalence between two configurations, we consider 4D $ D1 $ brane dynamics in the presence of vanishing NS-NS fluxes ( $ \mathcal{B}_2=0 $) and take into account the field redefinition of the following form,
\begin{eqnarray}
\tilde{\psi}=\Phi - \xi^0.
\label{en86}
\end{eqnarray}

Substituting (\ref{en86}) into (\ref{e38})-(\ref{e39}) we find,
\begin{eqnarray}
\cos\theta \dot{\theta}\dot{\Phi}-\dot{\theta}^2 \cos\theta + \ddot{\Phi}\cos\theta - \ddot{\theta}\sin\theta &=&0\\
\sin\theta \dot{\theta}\dot{\Phi}-\ddot{\Phi}\cos\theta &=&0
\end{eqnarray}
which precisely matches with TNC $ D0 $ brane dynamics (\ref{e20})-(\ref{e21}) subjected to the identification, $ \Lambda = \frac{1}{2 l^2_s} $. 
\section{Summary and final remarks}
We now summarise the key findings of the analysis. In the first part of the paper, we establish classical integrability of relativistic $ D1 $ branes (propagating over 4D relativistic background with a null isometry direction) following two traditional approaches - (i) the Kovacic's algorithm of showing classical (non)integrability and (ii) the standard formulation of Lax connections. Our analysis reveals that both of these approaches produce mutually convincing results.  It has been further shown that following a double null reduction of the world-volume coordinates, these $ D1 $ branes could be mapped into an \emph{integrable} Newton-Cartan $ D0 $ brane configuration those persisting over a 3D torsional Newton-Cartan geometry. Finally, we show that following a trivial field redefinition of the $ D1 $ brane world-volume fields, these two configurations could be mapped into each other in the presence of vanishing NS fluxes.  \\ \\ 
{\bf {Acknowledgements :}}
 The author is indebted to the authorities of IIT Roorkee for their unconditional support towards researches in basic sciences. \\\\ 

\end{document}